\documentclass[trackchanges]{aastex701}
\usepackage{multirow}
\usepackage{ulem}

\begin{document}

\title{Dark-Matter-Deficient Galaxies from Collisions: A New Probe of Bursty Feedback and Dark Matter Physics}

\author[0000-0003-1215-6443]{Yi-Ying Wang}
\affiliation{Key Laboratory of Dark Matter and Space Astronomy, Purple Mountain Observatory, Chinese Academy of Sciences, Nanjing 210033, People's Republic of China}
\email{}

\author[0000-0002-5421-3138]{Daneng Yang}
\affiliation{Key Laboratory of Dark Matter and Space Astronomy, Purple Mountain Observatory, Chinese Academy of Sciences, Nanjing 210033, People's Republic of China}
\affiliation{School of Astronomy and Space Science, University of Science and Technology of China, Hefei, Anhui 230026, People's Republic of China}
\email[show]{yangdn@pmo.ac.cn}

\author{Keyu Lu}
\affiliation{Key Laboratory of Dark Matter and Space Astronomy, Purple Mountain Observatory, Chinese Academy of Sciences, Nanjing 210033, People's Republic of China}
\email{}

\author[0000-0002-7275-8561]{Yue-Lin Sming Tsai}
\affiliation{Key Laboratory of Dark Matter and Space Astronomy, Purple Mountain Observatory, Chinese Academy of Sciences, Nanjing 210033, People's Republic of China}
\affiliation{School of Astronomy and Space Science, University of Science and Technology of China, Hefei, Anhui 230026, People's Republic of China}
\email{}

\author[0000-0002-8966-6911]{Yi-Zhong Fan}
\affiliation{Key Laboratory of Dark Matter and Space Astronomy, Purple Mountain Observatory, Chinese Academy of Sciences, Nanjing 210033, People's Republic of China}
\affiliation{School of Astronomy and Space Science, University of Science and Technology of China, Hefei, Anhui 230026, People's Republic of China}
\email[show]{yzfan@pmo.ac.cn}

\begin{abstract}
High-velocity collisions between gas-rich ultra-diffuse galaxies present a promising formation channel for dark-matter-deficient galaxies (DMDGs). Using hydrodynamical simulations, we show that the progenitors' baryonic binding energy, $|E_{\rm bind}|$, critically controls the outcome. Repeated potential fluctuations, e.g., from bursty feedback, inject energy and reduce $|E_{\rm bind}|$ by $\approx 15\%$, yielding fewer but substantially more massive DMDGs. By contrast, elastic self-interacting dark matter (SIDM) produces comparable cores without lowering $|E_{\rm bind}|$, perturbing DMDG masses without clear enhancement. This differs from what happens in host halos, where SIDM-induced cores enhance dark matter tidal stripping while keeping baryons compact and resilient to tidal effects. The contrasting roles of SIDM may provide a means to distinguish feedback-formed halo cores from those created by SIDM. Among 15 paired simulation runs, 13 show higher DMDG masses in the weakened-binding case, and about two thirds exhibit $>100\%$ mass enhancements. The simulations also predict systematically lower gas fractions due to sustained post-collision star formation, yielding a clean observational signature. Upcoming wide-field imaging (CSST, LSST), HI surveys (FAST), and kinematic follow-up will be crucial to test this scenario.
\end{abstract}

\section{Introduction} 
Within the standard cosmological model, the cosmic baryon fraction is only one-fifth that of the dark matter, and galaxies are expected to reside in more massive dark matter halos. 
As a result of galaxy formation governed by complex processes, the dark matter content of typical dwarf galaxies is expected to exceed their stellar mass by roughly two orders of magnitude~\citep{1986ApJ...303...39D,Hopkins:2013vha,Vale:2004yt,2013ApJ...770...57B}.  

In this context, the observation of galaxies with exceptionally low dark matter content challenges the $\Lambda$CDM framework~\citep{2018Natur.555..629V,2019ApJ...874L...5V,Guo:2019wgb,PinaMancera:2021wpc,2023A&A...675A.143C}. A notable case is NGC~1052–DF2 and DF4 \citep{2018Natur.555..629V,2019ApJ...874L...5V}, commonly described as dark-matter-deficient galaxies (DMDGs). Both of these are ultra-diffuse systems, with velocity dispersion measurements consistent with little or no dark matter. 
Proposed formation scenarios include tidal stripping in cored halos \citep{2018MNRAS.480L.106O,yang:2020iya,2020MNRAS.494.1848S,Zhang:2024qem,Zhang:2024qmh} and high-velocity collisions that produce dark-matter-deficient remnants \citep{2022Natur.605..435V,2024ApJ...966...72L,Keim:2025gfn}. 
Notably, \citet{Guo:2019wgb} identified $19$ DMDGs from a sample of $324$ SDSS dwarfs, with $14$ located well beyond the virial radii of nearby groups or clusters, where environmental stripping and recent interactions are unlikely. 

Existing cosmological simulation studies have identified DMDGs as tidally stripped satellites~\citep{2022NatAs...6..496M,2019MNRAS.488.3298J}. 
In this scenario, controlled simulations further map the orbital conditions and galaxy-halo configurations required to form such systems, offering a way to test models of elastic self-interacting dark matter (SIDM) that produce cored halos without making baryons too diffuse~\citep{yang:2020iya}.
However, the identification of DMDGs in isolation remains challenging, hindered by the competing requirements for both high resolution and a large field of view.
Recently, \citet{2024ApJ...966...72L} used the TNG100-1 simulation~\citep{2019ComAC...6....2N} to search for gas-rich dwarf pairs whose orbital parameters satisfy requirements for producing DMDGs near massive hosts, finding $\sim 10$ such collisions in a $(100\ {\rm Mpc})^3$ volume over $z\simeq 3\rightarrow 0$. 
The mass and spatial resolution of TNG100 are insufficient to follow the hydrodynamics required to produce such systems, but the implied event rate broadly agrees with the handful of field DMDGs reported by \citet{Guo:2019wgb}, suggesting that dwarf-dwarf collisions provide a cosmologically plausible channel for generating observable DMDGs. 
Though how collision velocity, disk structure, and initial configuration affect the DMDG formation has been examined in some recent simulations \citep{2020ApJ...899...25S,2024ApJ...966...72L},
it remains unclear what physical quantity critically controls the mass and yield of DMDGs.  

In this work, we identify the baryonic binding energy of the progenitors, $|E_{\rm bind}|$, as the relevant control parameter, directly linking energy injection in dwarf halos to DMDG yields. 
We employ controlled hydrodynamical simulations of gas-rich ultra-diffuse galaxy (UDG) collisions to assess the potential of these systems as probes of energy injection in dwarf halos, positioning DMDG observations a new observational window for testing the underlying physics. As both baryonic feedback~\citep{Navarro:1996bv,2006Natur.442..539M,2010Natur.463..203G,Burger:2022jid} and self-interacting dark matter (SIDM)~\citep{Spergel:1999mh,2018PhR...730....1T} have been proposed as mechanisms for generating cores in halos, we also set up simulations to test the capabilities of feedback and SIDM in boosting DMDG formations. Our results reveal distinct differences in their effects on DMDG formation. As a result, these two core formation mechanisms are expected to produce different DMDG populations in isolated and host environments. Measuring these populations offers an observational method to advance understanding of galaxy formation and dark sector physics.

\section{Weakening gravitational binding by energy injection.}

To unbind baryons from halos more efficiently, we need processes that reduce the gravitational binding of the entire system. 
Although energy injection into halos can weaken the binding of their baryonic component, most baryonic processes, such as gas cooling, release their energy radiatively, thereby dissipating rather than adding energy to the gravitationally bound system. 
In contrast, repeated potential fluctuations can heat dark matter in the inner regions, reducing the gravitational binding after relaxation and flattening the inner halo density profile.
Such fluctuations arise naturally from bursty stellar feedback in dwarf galaxies~\citep{2012MNRAS.421.3464P,2018MNRAS.478..906C,2020MNRAS.491.4523F,LiZZ23mn}.
In novel dark matter models, recurrent collapses of solitonic cores can trigger bosenova-like outbursts that drive similar fluctuations~\citep{Fox:2023xgx,Koo:2025qac}.
Scattering among dark matter particles may also be exothermic, imparting kinetic energy to the final-state particles while keeping them bound to the halo~\citep{Vogelsberger:2018bok,Medvedev:2013vsa}.

\begin{figure*}[htbp!]
\centering
\includegraphics[width=\textwidth]{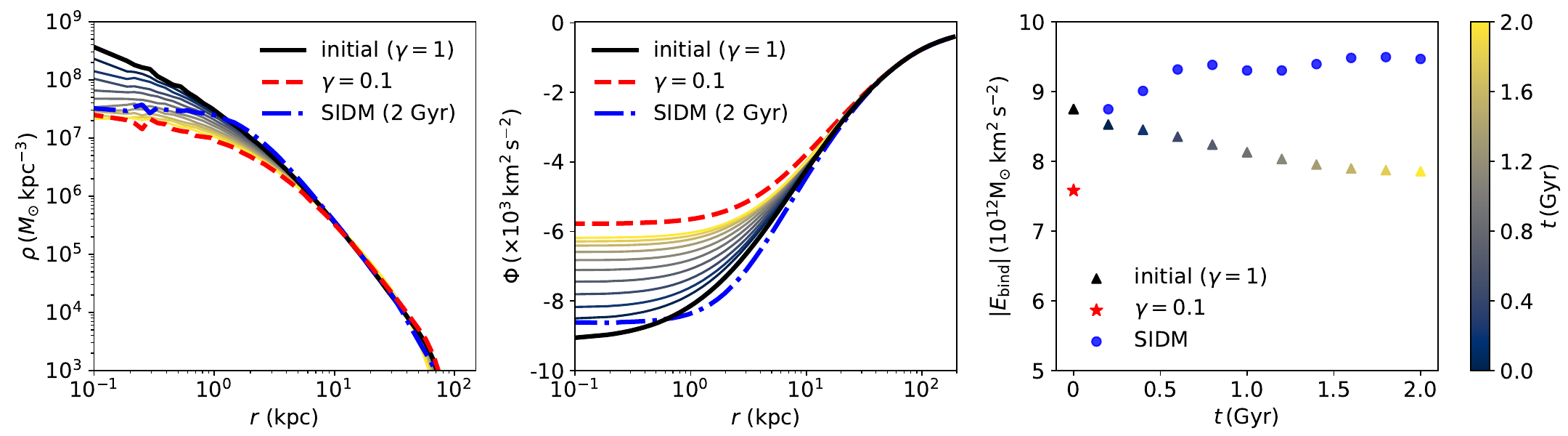}
\caption{\small 
Effect of energy injection into dark matter halos.
Repeated potential variations, such as those induced by bursty feedback, can flatten the inner halo densities (left), shallow the potential (middle), and decrease the absolute baryonic binding energy, $|E_{\rm bind}|$ (right).
The final snapshot resembles a $\gamma=0.1$ (red) profile, which features an inner density core and a reduced $|E_{\rm bind}|$ relative to the initial cuspy $\gamma=1$ profile (black), and we adopt this in our simulations. 
For comparison, we overlay SIDM results (blue) from a simulation with the same initial $\gamma=1$ condition and a cross section per mass $\sigma/m = 20~\rm cm^2/g$. 
The initial conditions correspond to the 1a and 1b benchmarks listed in \autoref{Tab:1}. 
}
\label{fig:Feedback}
\end{figure*}

As a quantitative illustration, we model the evolution of a halo density profile under energy injection, following the method of \citet{2012MNRAS.421.3464P}, which has been widely applied in various contexts~\citep{2012MNRAS.421.3464P,2014MNRAS.437..415D,2017MNRAS.472.2153P,2024PhRvD.110h3004A}. For demonstration, we present results for a spherical, isotropic NFW halo with a mass of $1.5\times 10^{10}~\rm M_{\odot}$ and a concentration parameter of $c=14$, corresponding to the 1a and 1b benchmarks in \autoref{Tab:1}. For each dark matter particle, we compute the total energy $E_0$ and specific angular momentum $j$.
A sequence of impulsive potential fluctuations is modeled by rescaling the potential by a small fraction, $\Delta V/V_0=0.14$, at each iteration. The energy of each particle is updated using the phase-averaged kick at fixed $j$
\begin{equation}
\Delta E=\frac{2nE_0}{(2+n)^2}\,\bigg(\frac{\Delta V}{V_0}\bigg)^2,
\end{equation}
where $n=2$ represents a harmonic oscillator potential.
After each impulse, we advance the simulation time by $0.2~$Gyr and reconstruct the density profile by summing the orbit-averaged radial probability distributions
\begin{eqnarray}
p(r;E,j)&\propto& \frac{1}{\sqrt{E-V_{\rm eff}(r,j)}}, \\
V_{\rm eff}(r,j)&=&V(r)+\frac{j^{2}}{2r^{2}}, 
\label{eq:prob}
\end{eqnarray}
over all dark matter particles. The baryon content is assumed to be a gas disk which takes spherically averaged pseudo-isothermal profile, $\rho_{\rm gas} = \rho_0 \left(1 + (r/r_g)^2 \right)^{-1}$, where $r_g = 2~\rm kpc$ and $\rho_0 = 2.46\times 10^6~\rm M_{\odot} ~ \rm kpc^{-3}$~\citep{2020MNRAS.491.4993K,2024ApJ...966...72L}. The disk height is set to $0.2~\rm kpc$.

To illustrate the impact of energy injection into halos in a model-independent manner, we adopt the following NFW-like extension for the halo density profile~\citep{Graham:2005xx,2018MNRAS.480L.106O}
\begin{equation}
\rho(r) = \frac{\rho_s}{\left(\frac{r}{r_s}\right)^{\gamma} \left( 1 + \frac{r}{r_s} \right)^{3 - \gamma}},
\end{equation}
where the central density slope is governed by $\gamma$, with smaller values of $\gamma$ corresponding to less gravitationally bound halos. \autoref{fig:Feedback} shows the evolution of the density profile (left), gravitational potential (middle), and $|E_{\rm bind}|$ (right) in 2 Gyr. 
We compute $E_{\rm bind}$ by integrating the gas density over the total gravitational potential ($\Psi_{\rm tot}$), 
$E_{\rm bind} = 4\pi\int dr r^2 \Psi_{\rm tot} \rho_{\rm gas}$. 
The results illustrate that impulsive feedback heats the inner orbits, leading to an expansion of centrally cored density profile, a shallowing of the central potential, and a reduction in the central baryonic binding energy. 
By $2$ Gyr, $|E_{\rm bind}|$ has decreased by over 10\%. 
We will use collision simulations to show that even this modest reduction can substantially enhance the DMDG formation.

\begin{deluxetable*}{c|ccccccc|ccc}
\tablecaption{\label{Tab:1}%
Galaxy collision simulations with \texttt{Gadget-4}: initial collision configurations and properties of the resulting dark-matter-deficient galaxies (DMDGs).}
\tabletypesize{\footnotesize}
\tablewidth{40mm}  
\tablehead{
\colhead{BM} &\colhead{$M_{200, \, \rm DM}$} & \colhead{$M_{\rm gas}$} & \colhead{$v_r$} & \colhead{$R_s$} & \colhead{$c$} & \colhead{$\gamma$} & \colhead{$|E_{\rm bind}|$} & \colhead{$n$} & \colhead{$M_{\rm star}$} & \colhead{$M_{\rm gas}$} \\
\colhead{} &\colhead{($10^{10}\,M_\odot$)} & \colhead{($10^{10}\,M_\odot$)} & \colhead{(km s$^{-1}$)} & \colhead{(kpc)} & \colhead{} & \colhead{}  & \colhead{$E_0$} & \colhead{} & \colhead{($10^{8}\,M_\odot$)} & \colhead{($10^{8}\,M_\odot$)}
}
\startdata
1a & \multirow{3}*{1.5} &\multirow{3}*{0.15} &\multirow{3}*{$400$} &\multirow{3}*{$2$} &\multirow{3}*{$14$} & 1  & 8.75 & 1 & 0.38 & 1.49 \\
1b & &      &     &   &    & 0.1& 7.58 & 1 & 8.61 & 4.85 \\
1c & &      &     &   &    & SIDM & 9.47 & 1 & 0.46 & 2.97 \\
\hline
2a &\multirow{3}*{2} &\multirow{3}*{0.2} &\multirow{3}*{$400$} &\multirow{3}*{$2$} &\multirow{3}*{$14$} & 1   & 14.73 & 4 & 0.10 & 2.13 \\
2b & &     &     &   &    & 0.1 & 12.49 & 1 & 6.09 & 6.72 \\
2c & &     &     &   &    & SIDM & 15.92 & 2 & 0.69 & 3.70 \\
\hline
3a &\multirow{3}*{2} &\multirow{3}*{0.2} &\multirow{3}*{$280$} &\multirow{3}*{$2$} &\multirow{3}*{$14$} & 1   & 14.73 & 2 & 4.89 & 3.70 \\
3b & &     &     &   &    & 0.1 & 12.49 & 1 & 16.80 & 4.97 \\
3c & &     &     &   &    & SIDM & 15.92 & 3 & 2.68 & 2.98 \\
\hline
4a &\multirow{3}*{1.0} &\multirow{3}*{0.1} &\multirow{3}*{$400$} &\multirow{3}*{$1.5$} &\multirow{3}*{$4$}   & 1   & 3.53 & 1 & 0.15 & 4.13 \\
4b & &     &     &     &    & 0.1 & 2.87 & 1 & 0.30 & 5.54 \\
4c & &     &     &     &    & SIDM &4.06  & 1 & 0.04 & 1.39 \\
\hline
5a &\multirow{3}*{1.5} &\multirow{3}*{0.15} &\multirow{3}*{$400$} &\multirow{3}*{$2$} &\multirow{3}*{$4$}& 1   & 7.17 & 1 & 6.96 & 8.32 \\
5b & & & & & & 0.1 & 5.82 & 1 & 9.61 & 5.05 \\
5c & & & & & & SIDM &8.28  & 1  & 3.74 & 4.24 \\
\hline
6a &\multirow{3}*{1.5} &\multirow{3}*{0.22} &\multirow{3}*{$400$} &\multirow{3}*{$2$} &\multirow{3}*{$4$} & 1   & 10.99 & 2 & 16.84 & 6.02 \\
6b & &      &     &   &   & 0.1 & 9.28 & 1 & 20.70 & 7.03 \\
6c & &      &     &   &   & SIDM &12.64  & 2 & 4.08 & 3.79 \\
\hline
7a &\multirow{3}*{1.5} &\multirow{3}*{0.22} &\multirow{3}*{$450$} &\multirow{3}*{$2$} &\multirow{3}*{$4$} & 1   & 10.99 & 2 & 2.39 & 7.74 \\
7b & &      &     &   &   & 0.1 & 9.28 & 2 & 3.23 & 3.96 \\
7c & &      &     &   &   & SIDM & 12.64 & 1 & 0.18 & 0.68 \\
\hline
8a &\multirow{3}*{1.5} &\multirow{3}*{0.15} &\multirow{3}*{$400$} &\multirow{3}*{$2$} &\multirow{3}*{$7$} & 1   & 7.87 & 3 & 1.62 & 3.38 \\
8b & &      &     &   &   & 0.1 & 6.47 & 1 & 11.15 & 3.91 \\
8c & &      &     &   &   & SIDM &8.74  & 3 & 0.25 & 3.33 \\
\hline
9a &\multirow{3}*{1.5} &\multirow{3}*{0.22} &\multirow{3}*{$400$} &\multirow{3}*{$2$} &\multirow{3}*{$7$} & 1   & 11.93 & 3 & 4.66 & 4.42 \\
9b & &      &     &   &   & 0.1 & 10.21 & 2 & 18.33 & 5.57 \\
9c & &      &     &   &   & SIDM & 13.18 & 1 & 15.79 & 3.60 \\
\hline
10a &\multirow{3}*{1.5} &\multirow{3}*{0.22} &\multirow{3}*{$280$} &\multirow{3}*{$2$} &\multirow{3}*{$7$} & 1   & 11.93 & 2 & 12.71 & 3.15 \\
10b & &      &     &   &   & 0.1 & 10.21 & 1 & 25.90 & 7.27 \\
10c & &      &     &   &   & SIDM & 13.18 &1  & 20.81 & 6.36 \\
\hline
11a &\multirow{3}*{1.5} &\multirow{3}*{0.15} &\multirow{3}*{$280$} &\multirow{3}*{$2$} &\multirow{3}*{$14$} & 1  & 8.75 & 1 & 0.22 & 2.14 \\
11b & &      &     &   &    & 0.1& 7.58 & 1 & 8.38 & 5.20 \\
11c & &      &     &   &    & SIDM &9.47  &3  & 3.50 & 3.14 \\
\hline
12a &\multirow{3}*{1.5} &\multirow{3}*{0.22} &\multirow{3}*{$400$} &\multirow{3}*{$2$} &\multirow{3}*{$14$} & 1  & 13.56 & 3 & 3.12 & 5.88 \\
12b & &      &     &   &    & 0.1& 11.68 & 1 & 13.15 & 5.23 \\
12c & &      &     &   &    & SIDM & 14.53 & 2 & 4.90 & 5.32 \\
\hline
13a &\multirow{3}*{2} &\multirow{3}*{0.2} &\multirow{3}*{$400$} &\multirow{3}*{$2$} &\multirow{3}*{$4$} & 1    & 11.53 & 1 & 14.64 & 6.60 \\
13b & &     &     &   &   & 0.1  & 9.75 & 1 & 19.49 & 5.88 \\
13c & &     &     &   &   & SIDM &13.20  &2  &7.60  & 4.26 \\
\hline
14a &\multirow{3}*{2} &\multirow{3}*{0.2} &\multirow{3}*{$400$} &\multirow{3}*{$2$} &\multirow{3}*{$7$} & 1    & 12.81 & 2 & 0.78 & 3.01 \\
14b & &     &     &   &   & 0.1  & 10.83 & 1 & 13.97 & 5.53 \\
14c & &     &     &   &   & SIDM  & 14.18 &1  & 11.60 & 5.19\\
\hline
15a &\multirow{3}*{5} &\multirow{3}*{0.5} &\multirow{3}*{$500$} &\multirow{3}*{$3.5$} &\multirow{3}*{$14$} & 1 & 78.10 & 3 & 13.81 & 4.98 \\
15b & &     &     &     &    &0.1 &66.40 & 1 & 26.08 & 5.11 \\
15c & &     &     &     &    &SIDM &84.24  &4  &11.38  &4.89  \\
\enddata
\tablecomments{The properties listed are: (1) dark matter halo mass $M_{200}$, (2) total gas mass, (3) initial relative collision velocity $v_r$, (4) gaseous disk scale radius, (5) concentration $c$, (6) $\gamma$, (7) the absolute value of the baryonic binding energy $|E_{\rm bind}|$, presented in unit $E_0\equiv (10^{12}\,M_\odot\,\rm km^{2} \, s^{-2})$, (8) number of resulting DMDGs with $M_{b}>10^6 M_\odot$, (9) stellar mass of the most massive DMDG, and (10) its gas mass. All properties are measured within $10\,{\rm kpc}$ at $t=2.0\,{\rm Gyr}$.}
\end{deluxetable*}

We also contrast this with the effect of SIDM, a distinct core formation mechanism. To this end, we simulate the same initial condition as in the 1a and 1b benchmarks (\autoref{Tab:1}), considering only gravity and using the \texttt{Gadget-2}-based SIDM module developed in \citet{yang220503392} and \citet{yang:2022mxl}, adopting a cross section of $\sigma/m = 20~\rm{cm^2 g^{-1}}$. In \autoref{fig:Feedback}, the evolved density profile at $2~\rm Gyr$ is shown in gray. While the SIDM halo develops a core of comparable size to that formed by feedback, its gravitational potential remains deep, and $|E_{\rm bind}|$ slightly increases rather than decreases. This suggests that baryonic binding, if quantified observationally, could serve as a diagnostic to distinguish between the two core formation mechanisms.

\autoref{Tab:1} summarizes the configurations and results of representative simulations. 
The labels $\gamma=0.1$ and $\gamma=1$ correspond to simulations with and without weakened baryonic gravitational binding. 
We model the density profiles using an extended NFW profile with an inner slope controlled by the parameter $\gamma$. 

\section{Collision Simulation Setup}
As described above, we construct two contrasting sets of initial conditions, one with $\gamma = 1$ (cuspy) and the other with $\gamma = 0.1$ (cored, with weaker binding), while keeping the total halo mass and concentration fixed. The detailed simulation parameters are summarized in \autoref{Tab:1}. We adopt a particle mass of $1.55\times10^4 \, \rm M_{\odot}$ \footnote{For the final sample with $M_{200 , \rm DM} = 5\times10^{10} \, M_{\odot}$, the particle mass is $3.95\times10^4 \, \rm M_{\odot}$.} and a gravitational softening length of $\epsilon=0.2 \rm \, kpc$ that satisfies $\epsilon \approx \frac{4r_{200}}{\sqrt{N_{200}}}$ \citep{2003MNRAS.338...14P}. The simulated halos span masses of $1-2 \times 10^{10}~\rm M_{\odot}$ and concentrations in the range $4-14$. 
Assuming these halos host gas-rich UDGs, we model the baryonic content with a gas-only disk, as described previously, and vary the disk height between $0.15$ and $0.35~\rm kpc$. 
This maximizes the effectiveness of DMDG formation, as stars behave as a collisionless component during the collision and cannot be efficiently expelled from the progenitors.
The gas-to-halo mass ratio is set to $0.1$ or $0.22$, with gas scale radii $r_g$ in the range $1.5-3.5$ kpc. 
These choices are motivated by observations of gas-rich UDGs, where internal turbulence and stellar feedback maintain gas densities below the threshold for molecular hydrogen formation, $n_g \lesssim 1~\mathrm{cm}^{-3} \approx 3 \times 10^7~ M_\odot\mathrm{kpc}^{-3}$, thereby suppressing star formation~\citep{DiCintio:2016ehs, 2018MNRAS.478..906C}.
The population and properties of gas-rich UDGs have attracted considerable attention and been extensively explored in recent studies~\citep{2017A&A...601L..10P,Jiang:2018iut,2018MNRAS.478..906C,2018A&A...614A..21J,ManceraPina:2020ujo,2022ApJ...936..166K}. 

Equilibrium initial conditions are generated using \texttt{DICE} \citep{2014A&A...562A...1P}, with the gas disk stabilized by enforcing a Toomre $Q_{\rm gas} > 1.5$~\citep{1964ApJ...139.1217T,2008gady.book.....B,2016ApJ...831....1W}. The gas is initialized at $T = 10^4$~K and metallicity $Z = 0.1 Z_\odot$. We evolve the collision simulations using \texttt{Gadget-4}~\citep{2021MNRAS.506.2871S}, incorporating radiative cooling of primordial gas and star formation based on the multiphase ISM model of \citet{Springel:2002uv}. 
The averaged thermodynamic impact of stellar feedback is modeled as an effective pressure that prevents the gas from collapsing too quickly. 
To obtain the SIDM prediction, we set up a hydrodynamical CDM simulation using progenitors taken from the isolated SIDM simulation (\texttt{Gadget-2}) snapshot at 2 Gyr. 
We verified that switching off SIDM after this initialization yields a stationary profile in isolation. 
Further details are provided in \autoref{app:3}.

Our collision simulations build upon findings in previous studies, where dependencies of simulation outcomes on the collision velocities, progenitor halo mass and concentration, gas distribution, and numerical resolution were systematically explored. 
We adopt simulation setups similar to those in \citet{2020ApJ...899...25S} and \citet{2024ApJ...966...72L}, with relative collision velocities ($v_r$) ranging from $300-600~\rm km/s$. Our primary choice of $v_r = 400~\rm km/s$ lies within the preferred range reported in the literature. We also perform simulations with $v_r = 280$, $450$ and $500~\rm km/s$.
The two colliding progenitors are initialized with identical conditions, placed at a separation of $60$ kpc along their relative velocity vector and with an impact parameter of $2$ kpc.
We focus on contrasting paired samples with $\gamma = 1$ and $\gamma = 0.1$, as well as an SIDM scenario, testing them under a range of conditions. 
All the collision simulations in this work use the same module for gas and stellar physics. It follows that all the simulation outcomes differ only because of the initial conditions.
The initial condition files for the simulations listed in \autoref{Tab:1} and the corresponding configuration files for {\tt Gadget-4} are publicly available at \url{https://pan.cstcloud.cn/s/xbilapRkS5Q}.

\begin{figure}[ht!]
\centering
\includegraphics[width=0.8\textwidth]{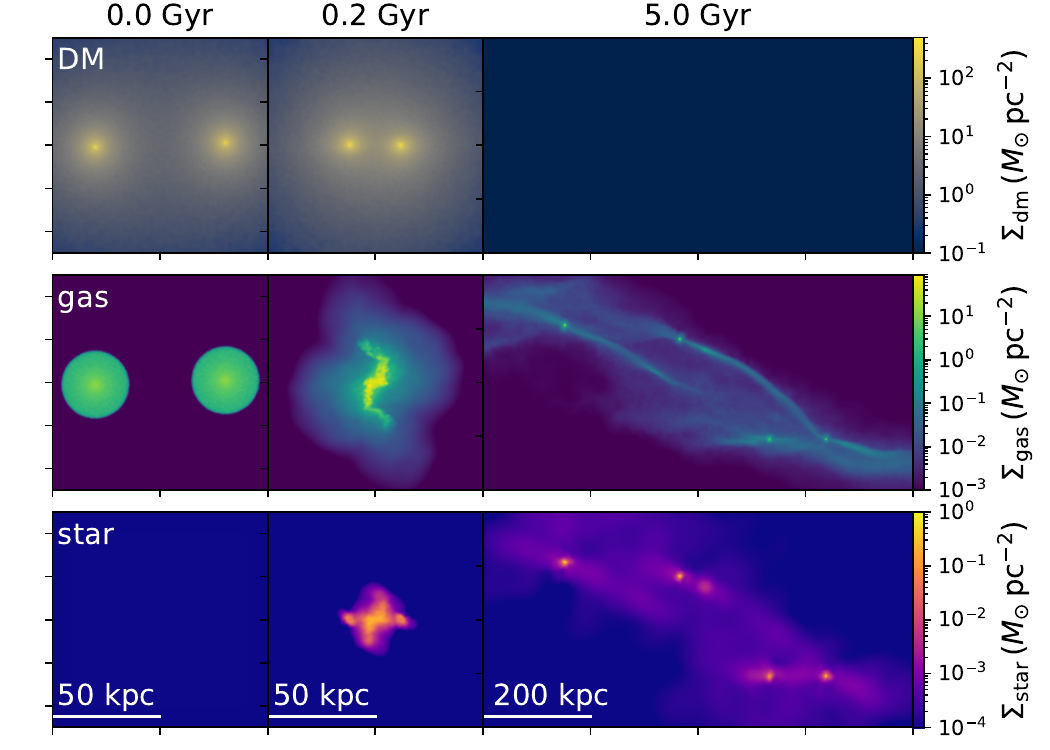}
\caption{\small 
Formation of DMDGs from the collision of gas-rich dwarf galaxies.
The initial setup (left) has two progenitors approaching at a close separation. 
They collide within 0.2 Gyr (middle), displacing gas from the halo centers and triggering efficient star formation.
By 5 Gyr (right), the expelled baryons have collapsed into several DMDGs that are far from their progenitors. 
The surface density distributions of dark matter (top), gas (middle), and stars (bottom) are presented at the corresponding snapshots for the 2b ($\gamma=1$) benchmark.
}
\label{fig:BKE15g01_collision}
\end{figure}

\section{Collisional formation of DMDGs.}

\autoref{fig:BKE15g01_collision} shows simulation snapshots of the surface densities of dark matter (top), gas (middle), and stars (bottom) for the benchmark 2a ($\gamma=1$) in \autoref{Tab:1}. Three stages of the collision are displayed. The first snapshot depicts the initial setup, with two progenitors approaching each other at close separation so that the simulation outcome primarily concerns the formation of DMDGs rather than the internal dynamics of the progenitors. The systems collide within 0.2 Gyr, displacing their gas from the halo centers and triggering efficient star formation. By 5 Gyr, the expelled baryons have already collapsed into several DMDGs and moved far from their progenitors. As shown in \autoref{Tab:1}, progenitors with $\gamma=0.1$ systematically produce more massive DMDGs. With $v_r=400~\rm km/s$, most of the progenitor gas is expelled. To understand the origin of these differences, we quantify the role of tidal effect and star formation efficiency in regulating gas separation and collapse during the collision.

\begin{figure*}[ht!]
\centering
\includegraphics[width=0.99\textwidth]{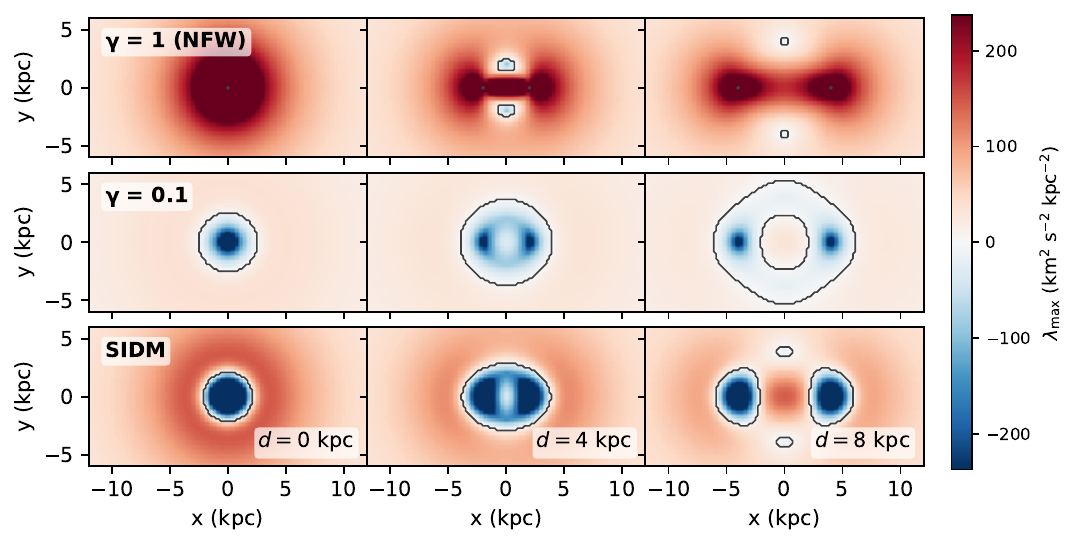}
\caption{Effect of tides on extended gas debris. The figure shows the largest eigenvalue ($\lambda_{\max}$) of the tidal tensor on the mid-plane for halo pairs with inner slopes of $\gamma=1$ (top), $\gamma=0.1$ (middle), and a cored SIDM profile (bottom), displayed at separations of $d=\{0,4,8\}~\rm kpc$ (columns).
Warm colors denote tidal stretching along at least one principal direction ($\lambda_{\max}>0$), while cold colors indicate fully compressive tides with all eigenvalues negative ($\lambda_{\max}<0$). The halo parameters follow the first benchmark listed in \autoref{Tab:1}.
}
\label{fig:tidal}
\end{figure*}

To quantify how the tide acts on an extended gas debris, we evaluate on the mid plane the largest eigenvalue ($\lambda_{\rm max}$) of the tidal tensor $T\equiv-\nabla\nabla\Phi$, for which negative eigenvalues imply convergent relative acceleration. A positive $\lambda_{\rm max}$ corresponds to stretching tides, which promote the separation and fragmentation of expelled gas into multiple low-mass condensations. \autoref{fig:tidal} shows the head-on collisions between two identical progenitor halos, using parameters from the first benchmark in \autoref{Tab:1}, at varying separations: $d=0$ (left), $4$ (middle), and $8~\rm kpc$ (right) for the $\gamma=1$ (top), $\gamma=0.1$ (middle), and SIDM (bottom) cases. 
We highlight the $\lambda_{\rm max}=0$ contour, within which all eigenvalues are negative and the tide is compressive. 
The $\gamma=0.1$ model develops broad, contiguous compressive regions around the origin across all separations, consistent with debris that readily coalesces into fewer, more massive clumps. A shallower inner slope renders the central potential more nearly harmonic, weakens shear, and suppresses stretching along the collision axis. 
By contrast, the cuspy $\gamma=1$ case exhibits only small, fragmented compressive islands around shear-dominated zones, consistent with the formation of multiple low-mass condensations. The SIDM core is encircled by a ring of stretching tides that inhibits gas confinement. The central compressive region still facilitates gas condensation, but the formation of DMDG is significantly suppressed compared with the $\gamma=0.1$ scenario.

As tidal forces are exerted at scales larger than the clumps to form, they operate as a ``wind'', which shallows or steepens the gravitational potential as it sweeps over a region that would collapse, depending on whether the tide is repulsive or compressive. 
Consequently, in the cuspy case ($\gamma=1$), the stronger repulsive tidal wind causes the expelled gas to fragment into several small, low-mass condensations. In contrast, in the weakened-binding case ($\gamma=0.1$), the debris remains in contact more easily, allowing it to coalesce into fewer but more massive DMDGs. 
In \autoref{app:2}, we provide further evidence showing that the tidal wind is sufficiently strong to affect the gas condensation right after the collision. 

In a weakened gravitational binding scenario, gas is expected to escape more readily during collisions. To illustrate this, we reduce the collision velocity of the 3a and 3b benchmarks to $280~\mathrm{km/s}$ (see \autoref{Tab:1} for setup and results). The escaping gas mass reaches $3.8 \times 10^9~M_{\odot}$ in the $\gamma = 0.1$ case, compared to $3.3 \times 10^9~M_{\odot}$ for $\gamma = 1$. Both the stellar and gas masses bound to the resulting DMDGs are substantially higher in the $\gamma = 0.1$ scenario, with enhancements of $240\%$ and $34\%$, respectively. The most massive DMDG formed in our simulations has a stellar mass of $2.60 \times 10^9~M_{\odot}$ in 15b, obtained in the $\gamma = 0.1$, $v_r = 500~\mathrm{km/s}$ run. These results demonstrate that reduced baryonic binding favors the formation of more massive DMDGs.

\begin{figure}[ht!]
\centering
\includegraphics[width=0.49\textwidth]{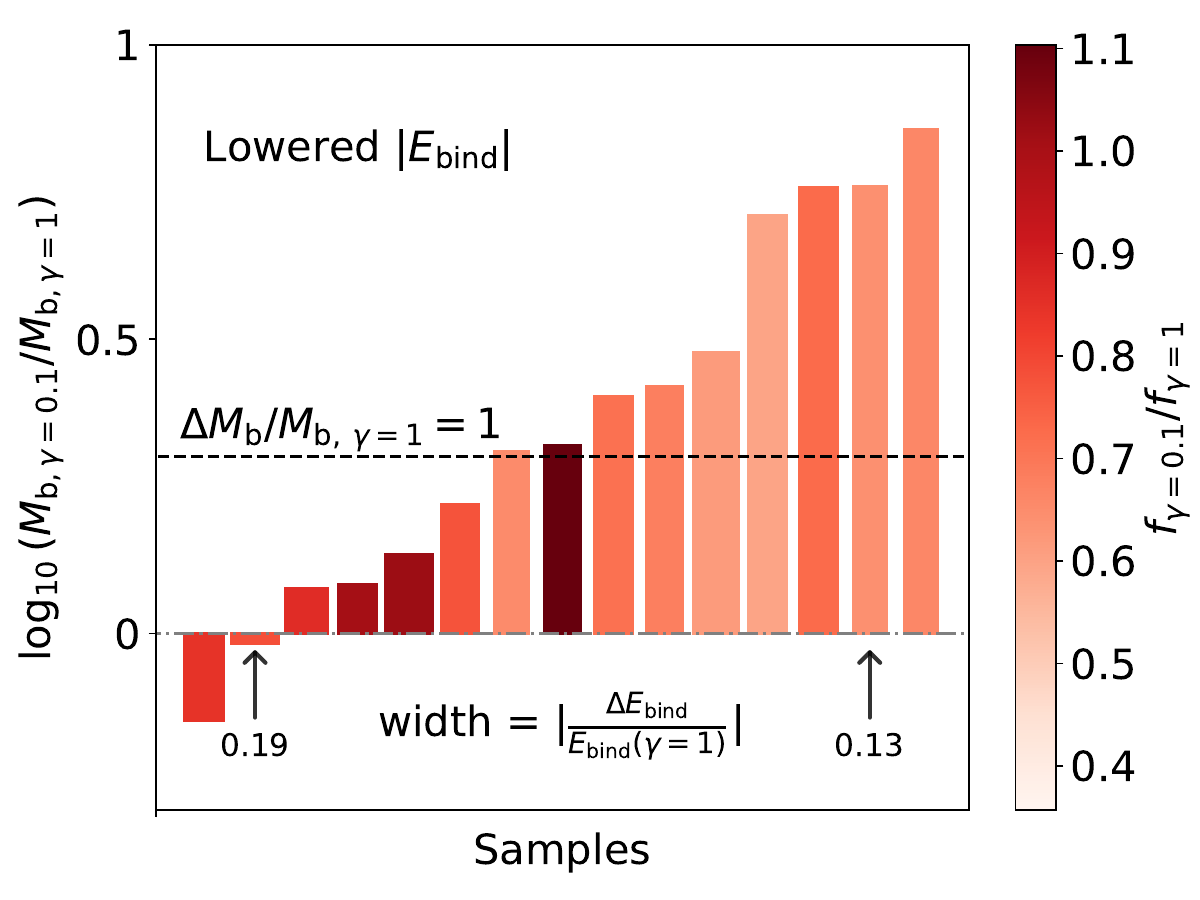}
\includegraphics[width=0.49\textwidth]{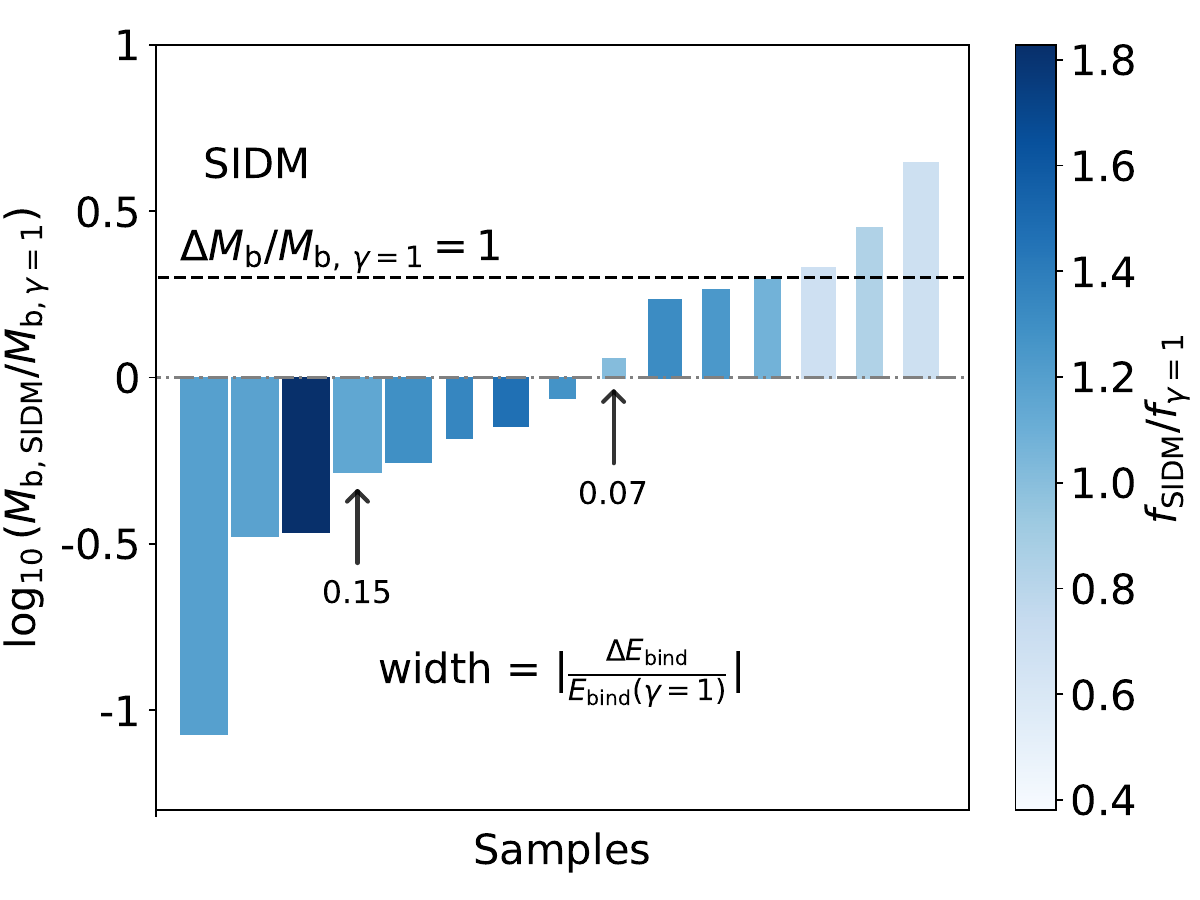}
\caption{Characteristics of the most massive DMDGs in 15 paired simulations. The bar chart summarizes the ratios of the most massive DMDG masses in the $\gamma = 0.1$ (left) and SIDM (right) scenarios, each normalized relative to the $\gamma = 1$ case. 
Bars are arranged in ascending order of the DMDG mass ratio, $M_{b,\gamma=0.1}/M_{b,\gamma=1}$ or $M_{b,SIDM}/M_{b,\gamma=1}$. The bar width scales with the relative change in binding energy, $|\Delta E_{{\rm bind}}/E_{{\rm bind},\gamma=1}|$, while the color indicates the gas fraction, $f = M_{\rm gas}/(M_{\rm gas}+M_{\rm stars})$. 
With lowered gravitational binding, the increase in the DMDG masses is predominantly positive. In 9 out of the 15 simulations, this increase exceeds 100\%, as reflected by the bars that rise above the line representing $\Delta M_b/M_{b,\gamma=1}=1$. 
In the SIDM scenario, mass increases and decreases occur with nearly equal frequency. There is also a weak anti-correlation between the magnitude of mass change and gas fraction.
}
\label{fig:bars}
\end{figure}

Aside from the examples shown, we simulated 15 sets of dwarf collisions with varied setups. \autoref{fig:bars} presents the most massive DMDG mass in the $\gamma = 0.1$ (left) or SIDM (right) cases, relative to the $\gamma = 1$ case, with bar widths proportional to the reduction in baryonic binding energy and colors indicating the gas fraction, $f = M_{\rm gas}/(M_{\rm gas}+M_{\rm stars})$. 
In the $\gamma=0.1$ cases on the left panel, 13 out of the 15 simulations yield an increase in DMDG mass, with 9 of these cases showing an enhancement greater than 100\%. 
The two exceptions occur for halos with extremely low concentrations ($c = 4$) and $v_r = 450~\mathrm{km/s}$, conditions that reduce the contrast between the two scenarios. 
On the right panel, we find that half of the most massive DMDGs in SIDM have lower masses than the $\gamma=1$ ones. The other half have increased masses with amplitudes close to those of the decreased ones, which are overall less significant than the increases in the $\gamma=0.1$ cases.  
These results agree with our expectations from the tidal wind analyses. 
Interestingly, the gas fractions in the $\gamma = 0.1$ runs are systematically lower than in their $\gamma = 1$ counterparts, demonstrating more sustained star formation following the collision in the $\gamma = 0.1$ cases. 
In SIDM, the gas fraction depicts an anti-correlation with the gas mass changes $\Delta M_b$, where the cases with increased DMDG masses have a lowered gas fraction, while the decreased mass cases have an increased gas fraction. 
In combination with the tidal analysis in \autoref{fig:tidal}, we find that the compressive (blue colors) and repulsive (red colors) tides may have caused the decrease and increase in the gas fractions. 
While both tides appear in SIDM, the compressive one dominates in the $\gamma=0.1$ case. 
Compressive tides can compress gas clumps, drive stronger star formation, and reduce gas fractions after a few Gyr. 
We provide details on the post-collisional star formation in \autoref{app:1}.
In addition, we present in \autoref{app:4} analogous results that include the total DMDG mass and the case with the gas cooling rate reduced by half. 
Our findings remain largely unchanged in these alternative scenarios. 

\section{DMDGs from tidal stripping in low-velocity collisions}

So far in this work, we have focused on high-velocity collisions in which DMDGs form outside progenitor halos. 
Our results show that a lower $E_{\rm bind}$ efficiently promotes DMDG formation, whereas SIDM cannot. 
Intriguingly, the formation of DMDGs has been explored in the context of SIDM, but for satellite galaxies~(\cite{yang:2020iya,Zhang:2024qem,Zeng:2024xty}). 
For example, \citet{yang:2020iya} shows that a DF2-like DMDG in the NGC1052 host halo can form more easily under SIDM, where the cored density profile boosts the tidal stripping of dark matter, making the satellite galaxy deficient in dark matter.  
Such analyses suggest that SIDM also increases DMDG production, but in a manner different from stellar feedback. Therefore, a population of DMDGs, both in isolation and in hosts, offers a novel window to differentiate between the two core formation mechanisms.

To explore how stellar feedback and SIDM-generated cores differ in producing DMDGs, here we follow the setup of \citet{yang:2020iya}, placing our progenitor halo systems on orbits evolving in a host potential. 
The host galaxy consists of a dark matter halo and a stellar component. The halo follows an NFW density profile with $M_{200}=1.1\times10^{13}~ {\rm M_{\odot}}$ and $r_s=80 ~ \rm kpc$, while the stellar component is modeled as a Hernquist profile~\citep{1990ApJ...356..359H} with scale density $\rho_h=1.1\times10^{10}~\rm M_{\odot}\, kpc^{-3}$ and scale radius $r_h=1.2 ~ \rm kpc$.
We set up three simulations for the $\gamma=1$, $\gamma=0.1$, and SIDM scenarios, taking the 2a, 2b, and 2c benchmarks from \autoref{Tab:1} as initial satellites and put them at the same orbital apocenter, $380~ \rm kpc$ from the host center, with a tangential velocity of $27 ~ \rm km ~ s^{-1}$. The simulations use {\tt Gadget-2} and treat all baryon particles as collisionless. To accurately model the tidal evolution, we consider $\gtrsim 10^6$ dark matter simulation particles and set the masses of baryon particles to be equal. The SIDM simulation adopts $\sigma/m=5~\rm cm^2/g$, a value widely adopted in the literature to address small-scale challenges. The initial core from the 2c benchmark can be interpreted as coming from pre-infall gravothermal evolution.

\begin{figure}[ht!]
\centering
\includegraphics[width=0.49\textwidth]{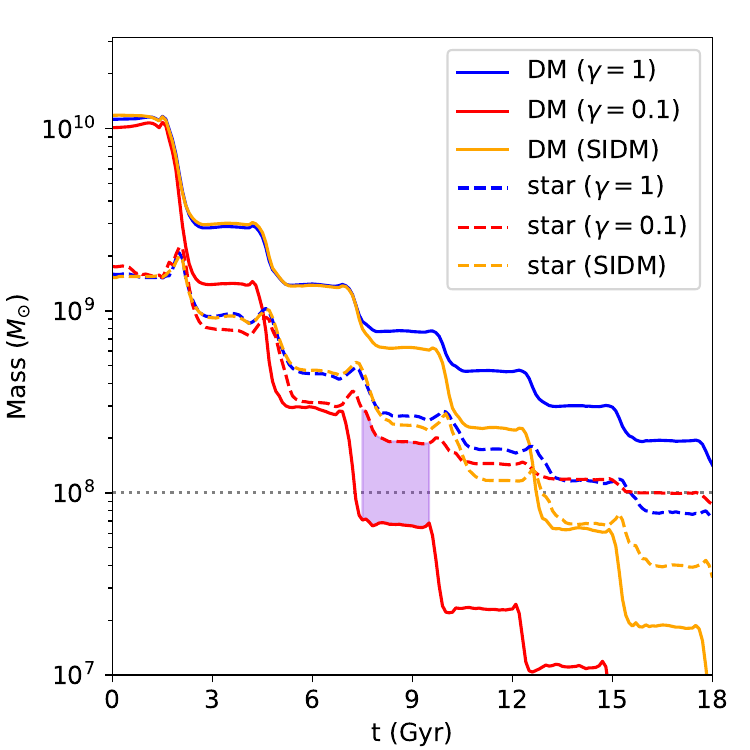}
\includegraphics[width=0.49\textwidth]{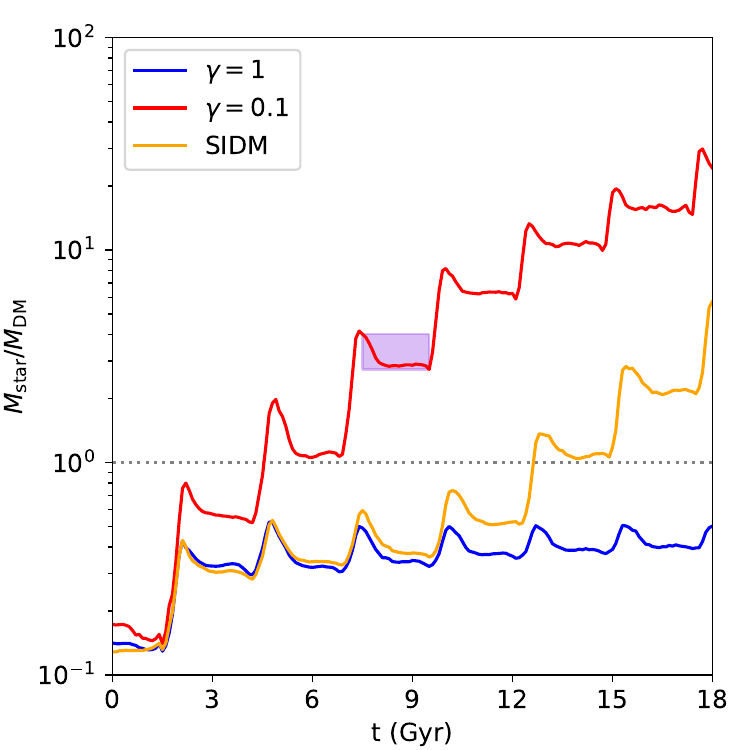}
\caption{ Tidal evolution of satellite systems in a $10^{13}~\rm M_{\rm \odot}$ host halo, with the initial conditions taken from the 2a, 2b, and 2c benchmarks from \autoref{Tab:1} for the $\gamma=1$ (blue), $\gamma=0.1$ (red), and SIDM (orange) scenarios. 
{\it Left:} Evolution of the bound dark matter (solid) and stellar (dashed) masses. 
The region giving rise to a close DF2 analog is shaded in light purple. 
{\it Right:} Evolution of the stellar-to-halo mass ratio. DMDGs are often selected by $M_{\rm star}/M_{\rm DM}>1$. The shaded area corresponds to the same region as in the left panel. }
\label{fig:TD}
\end{figure}

\autoref{fig:TD} presents the simulation results for the evolution of bound masses (left) and stellar-to-halo mass ratios (right). On the left panel, both the dark matter (solid) and the stellar (dashed) masses decrease along with the tidal evolution, with each steep mass loss corresponding to a pericentric passage. At the first of such passages, the $\gamma=0.1$ case shows already more significant mass losses in both dark matter and stars than the other two cases. This suggests that a lowered gravitational binding facilitates tidal stripping. The SIDM case is similar to the $\gamma=1$ case, since their potentials look similar beyond the core region. As the satellites’ tidal radii shrink along with their mass loss, tidal stripping removes masses from progressively inner regions, and we start to see that SIDM cores accelerate tidal stripping.  On the right panel, the stellar-to-halo mass ratio increases more rapidly in $\gamma=0.1$ right after the first pericentric passage, exceeding one in about 5 Gyrs. The growth continues and passes 10 around 12 Gyr, when the SIDM case starts to show a surge in the ratio above 1. 

Both the lowered gravitational binding and SIDM scenarios produce DMDGs within the age of the universe, but the required times and DMDG masses differ significantly. A DF2 analog of mass around $10^8~\rm M_{\odot}$ forms in bout 8 Gyr in the $\gamma=0.1$ case, and we shade the region of relevance in light purple. Such an analog appears in SIDM at around 13 Gyr, with a much lower stellar-to-halo mass ratio. These results reveal the different capabilities of stellar feedback and SIDM in producing DMDGs through tidal stripping. 

\section{Discussion and Conclusion}
Our controlled simulations show that energy injection weakens the progenitors' baryonic binding, elevating DMDG formation through collisions. Repeated potential fluctuations, as expected from bursty stellar feedback, may reduce $|E_{\rm bind}|$ by only 15\%, yet in two thirds of the simulations, the DMDG masses are enhanced by over 100\%. In contrast, SIDM creates cores without lowering $|E_{\rm bind}|$, resulting in heavier and lighter DMDGs from the $\gamma=1$ cases with similar chances.

In the regime of high mass ratios and low collision velocities, close encounters between satellite and host galaxies have been proposed as a primary channel for producing DMDGs within host systems~\citep{2022NatAs...6..496M,yang:2020iya,2019MNRAS.488.3298J,2018MNRAS.480L.106O}. We simulate the three scenarios under the same orbit and host halo, showing that both baryon feedback and SIDM can foster DMDG formation through enhanced tidal stripping, but with different capabilities. The relative abundance of DMDGs in the field versus those in host halos thus provides a potential observational handle to distinguish between core formation driven by baryonic feedback and that induced by SIDM. 

The distinct formation pathways of DMDGs may lead to contrasting observational signatures. Systems produced through tidal stripping resemble ordinary satellites but typically exhibit lower surface brightness. In contrast, those formed from post-collision tidal debris contain a substantial stellar component with nearly uniform ages and metallicities, created in a short burst at the time of the encounter. 
Another class of baryon-dominated systems is the tidal dwarf galaxy, which forms from tidal-tail gas ejected during a strong interaction involving a massive disk~\citep{Bournaud:2006qz,Gentile:2007gp,2012MNRAS.419...70K,2024A&A...689A.206Z}. These galaxies are typically embedded within tidal tails and generally lack bright globular clusters~\citep{2024ApJ...966...72L}.

Future observations of such systems will be highly informative. Wide-field surveys (LSST, WFST, CSST, Roman, Euclid) will significantly expand the catalogs of low-surface-brightness galaxies and enable the identification of DMDG candidates through their morphologies, stellar populations, and associated globular clusters. Crucially, 21-cm surveys (FAST, MeerKAT, VLA) will measure gas fractions and kinematics, confirming whether baryons alone can account for the observed dynamics. 

Extending this analysis to more realistic and cosmological settings will be essential for quantifying the DMDG population both in the field and within host environments. However, meeting the simultaneous demands of high resolution and large cosmological volume remains challenging. 
Current high-resolution simulations that resolve individual supernova explosions are restricted to controlled simulations, and cosmological simulations treat bursty feedback as subgrid physics~\citep{Zhang:2025xga,2024A&A...691A.231D,2022NatAs...6..496M,2019ComAC...6....2N}. Continued advances in simulations are therefore needed to clarify the formation processes and abundance of DMDGs.

\begin{acknowledgments}
We thank Fangzhou Jiang and Hui Li for useful discussions. 
This work is supported in part by the National Key Research and Development Program of China (No. 2022YFF0503304), the National Science Foundation of China (No. 12588101), the New Cornerstone Science Foundation through the XPLORER PRIZE, the China Manned Space Program (No. CMS-CSST-2025-A03), the Project for Young Scientists in Basic Research of the Chinese Academy of Sciences (No. YSBR-092), the Postdoctoral Fellowship Program of CPSF (No. GZB20250738), and the Jiangsu Funding Program for Excellent Postdoctoral Talent (No. 2025ZB209).
\end{acknowledgments}

\appendix \label{app}

\section{Stability of SIDM cores}\label{app:3}

\begin{figure}[ht!]
\centering
\includegraphics[width=0.49\textwidth]{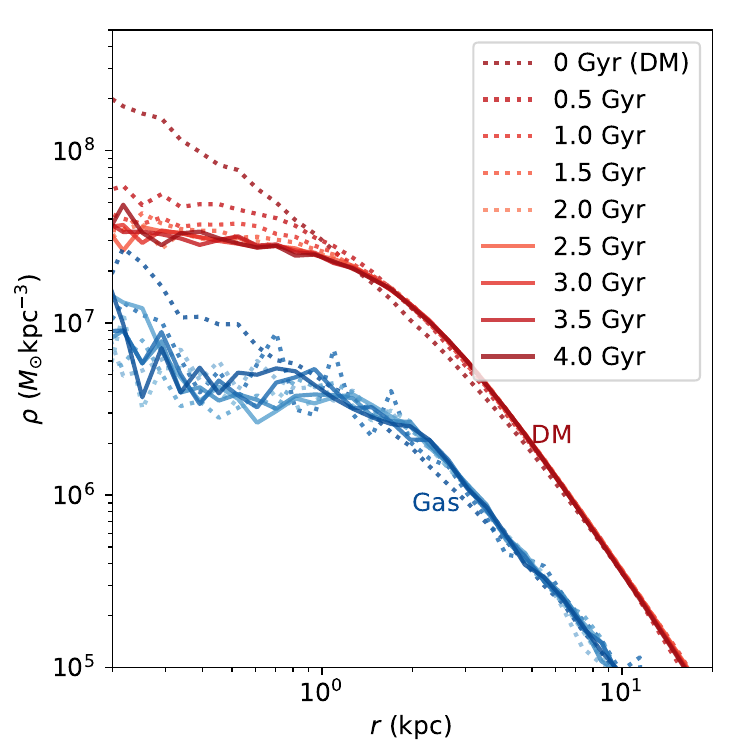}
\includegraphics[width=0.49\textwidth]{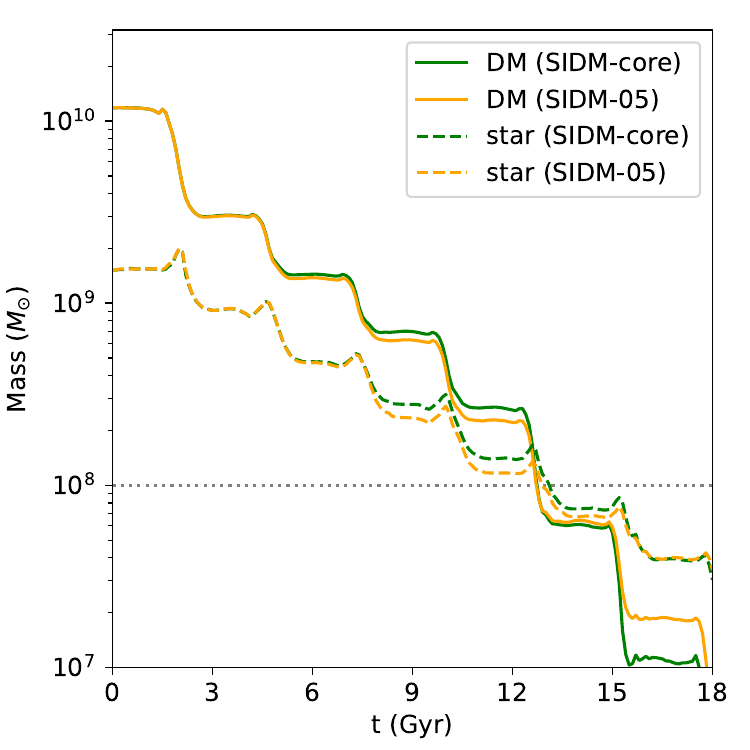}
\caption{\small 
{\it Left:} The stability test of SIDM halos. The progenitor system in the 1c simulation is first evolved under SIDM ($\sigma/m=20~\rm cm^2~g^{-1}$) for 2~Gyr (dotted lines), 
when a dark matter core rapidly forms and reaches a stable configuration.  
SIDM interactions are then turned off and the system is further evolved for 2~Gyr (solid lines). 
Both the dark matter (red) and stellar (blue) components remain stable, confirming that the SIDM-induced core is long-lived and justifying the neglect of SIDM in the subsequent collisional simulations. {\it Right:} The tidal tripping effect for the SIDM core phase with and without SIDM effect. The orange lines are consistent with those in \autoref{fig:TD}. The green lines presents the same initial condition to the orange lines, but do not include the SIDM effect during the simulation process.
}
\label{fig:SIDM_stable}
\end{figure}

The SIDM benchmark in our study is implemented by incorporating cored halos obtained from a pre-simulation.  
Specifically, we use {\tt Gadget-2} to evolve an initial system consisting of a gas disk and dark matter under gravity and SIDM ($\sigma/m=20~\rm cm^2~g^{-1}$) for 2~Gyr. 
The SIDM simulation is based on a well-tested module developed in prior works~\citep{yang220503392,yang:2022mxl}. Since {\tt Gadget-2} does not include radiative cooling and star formation modules, we pass the final snapshot from the pre-simulation to {\tt Gadget-4} for the hydrodynamical collision simulation. 
For computational efficiency, we adopt a relatively large SIDM interaction $\sigma/m=20~\rm cm^2~g^{-1}$, which leads to significant core formation within $\sim$ 2 Gyr. 
The resulting cored halo distribution can be interpreted as arising from lower cross sections with a longer evolution time ($t_{\rm evo}$), based on a degeneracy in which $t_{\rm evo}\sigma/m$ is roughly constant. For example, one can equivalently consider $\sigma/m=5~\rm cm^2~g^{-1}$ with an 8~Gyr pre-evolution.  

In our high-velocity dwarf collision simulations, the effect of SIDM primarily goes into the initial conditions. This is because the time scale of the dwarf collisions is very short. In the 2a benchmark, for example, the two halos pass each other and separate by about 100 kpc in just 0.4~Gyr. At this point, the region where DMDGs form has become baryon-dominated, and SIDM does not play a role. 
Aside from this, we provide additional justifications for our approach below. 

First, radiation pressure in the gaseous disk is subdominant compared to gravitational support; therefore, running the initial SIDM simulation without hydrodynamics does not break the equilibrium of the gas disk, allowing it to be transferred consistently into a subsequent hydrodynamical run.  
Second, the constant SIDM cross section per mass adopted here should be regarded as an effective constant cross section~\citep{yang220503392}.  
In velocity-dependent SIDM models consistent with observational constraints, the effect of SIDM during the collision can be neglected because the collision velocity is much larger than the typical particle velocities in the progenitor halos.  
Lastly, we assume that once SIDM cores form, the resulting cored profiles remain stable over relatively long timescales, even if SIDM interactions are subsequently turned off.  
This implies that SIDM halos reach a quasi-hydrostatic equilibrium, a point we explicitly verify with a toy simulation.  

The left panel in \autoref{fig:SIDM_stable} illustrates this verification for the progenitor system in the 1c simulation. During the first 2~Gyr, the halo evolves under SIDM: the dark matter core develops rapidly within the first $\sim$1~Gyr and then approaches a stable configuration. After 2~Gyr, SIDM is switched off and the simulation continues. Both the dark matter (red) and stellar (blue) components remain stable for at least another 2~Gyr.
These demonstrate that neglecting SIDM in the collisional simulation is a valid approximation.  

In our low-velocity collision simulations, where the satellite evolves in the host potential for over 10 Gyr, we have consistently simulated the SIDM effect. 
Here, we compare simulation results with and without SIDM enabled during evolution. 
\autoref{fig:SIDM_stable} shows the bound dark matter (solid) and stellar (dashed) mass evolution with (orange) and without (green) SIDM. 
As expected, their evolution trajectories are very close to each other. 
SIDM with a cross section per mass $\sigma/m=5~\rm cm^2/g$ only slightly boosts the tidal stripping, demonstrating that the SIDM effect primarily goes into the initial condition, even for these cored systems in the host.

\section{Tidal effect}\label{app:2}
\begin{figure*}[ht!]
\centering
\includegraphics[width=0.99\textwidth]{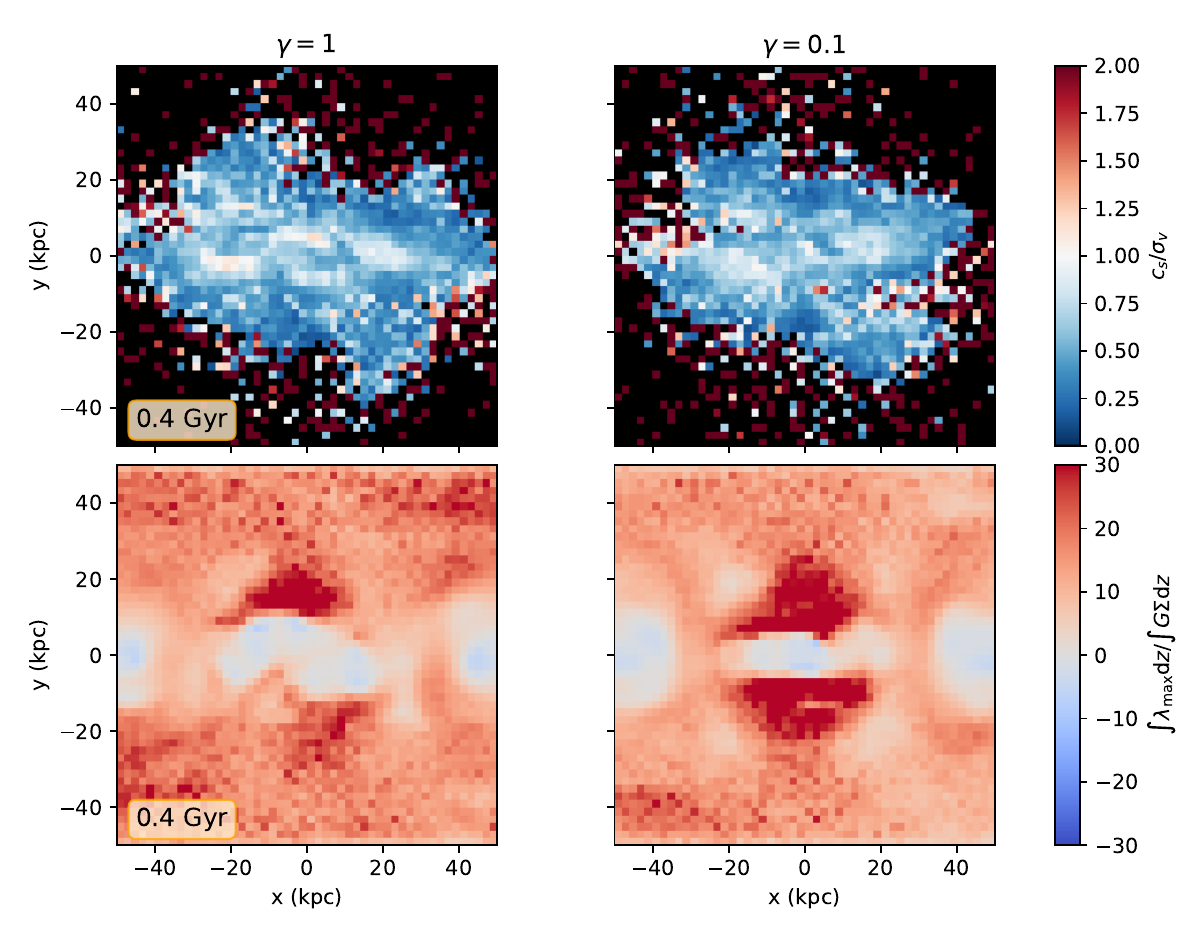}
\caption{ 
Additional contributing aspects of post collision gas condensation. 
The upper panels plot the ratio of the gas sound speed ($c_s$) to its velocity dispersion ($\sigma_v$). 
We analyze the 2a ($\gamma=1$) and 2b ($\gamma=0.1$) benchmarks from \autoref{Tab:1} and present results at $t=0.4~$Gyr, when DMDGs are about to form following the collision.
In the region where gas collapse, we find $\sigma_v\gg c_s$, which suggests gravity dominates over internal pressure in driving the dynamics of the gas.
The bottom panels illustrate an approximation of the tidal to self-gravitational acceleration ratio, defined as $a_{\rm tidal}/a_{\rm self}\sim \frac{\int \lambda_{\rm max} {\rm d}z}{ \int G\Sigma {\rm d}z}$. The results are presented for the same benchmarks as in the upper panels. 
A bulk of the region shows magnitudes greater than 1, implying that a stronger tidal effect shapes the gas collapse.  }
\label{fig:TD_wind}
\end{figure*}

The main text examined the impact of tidal forces on post-collision gas condensation. Here, we consider two additional factors, i.e., thermal pressure and self-gravity, and demonstrate that their effects are secondary at intermediate scales, where gas accumulates in regions likely to form DMDGs.

First, we show that thermal pressure ($\rho c_s^2$) is less significant than kinetic pressure ($\rho \sigma^2$). The upper panels in \autoref{fig:TD_wind} display maps of the sound speed to velocity dispersion ratio, $c_s/\sigma_v$, for the 2a and 2b benchmarks at $t=0.4~$Gyr. The results are consistently below one in regions of gas collapse, confirming that kinetic pressure dominates over thermal pressure.

To show that the tidal winds effectively reshape the gas collapse, we plot an approximate quatity for the ratio of tidal to self-gravitational acceleration, defined as $a_{\rm tidal}/a_{\rm self} \sim \frac{\int \lambda_{\rm max} {\rm d}z}{\int G\Sigma {\rm d}z}$, where $\lambda_{\rm max}$ is the largest eigenvalue of the tidal tensor. The lower panels of \autoref{fig:TD_wind} present the results considering the same benchmarks and snapshot as in the upper panels. The results show $|a_{\rm tidal}/a_{\rm self}|>1$ in most areas, suggesting that there is a phase of tidal dominance before DMDGs form under the baryon’s self-gravity. 

\section{Star formation in gas remnants}\label{app:1}

\begin{figure*}[ht!]
\centering
\includegraphics[width=0.99\textwidth]{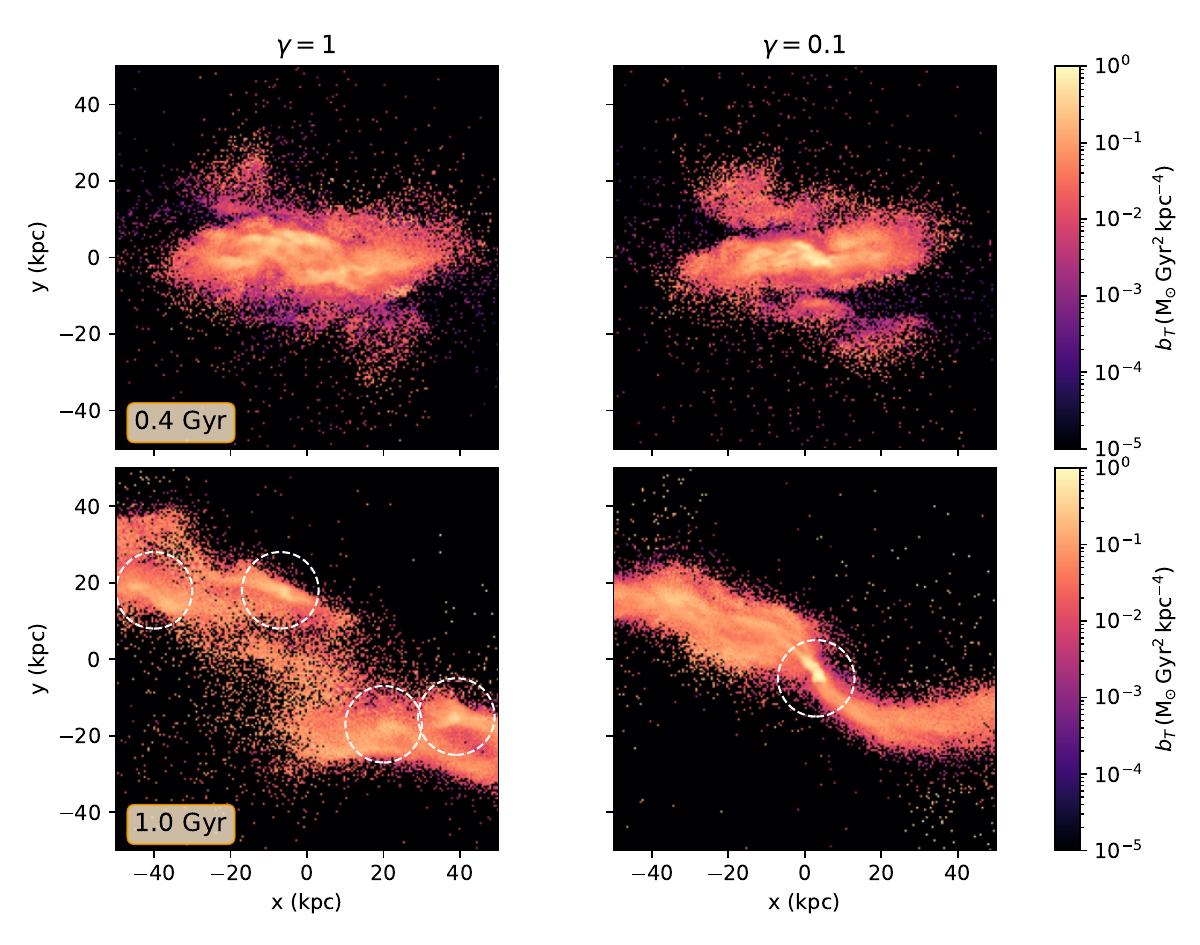}
\caption{\small 
Comparison of star formation efficiency in cored and cuspy benchmarks. 
Maps of the $b_T$, which traces the star formation efficiency, are shown for the cuspy ($\gamma=1$, 2a, left) and cored ($\gamma=0.1$, 2b, right) benchmarks at $t=0.4$ Gyr (top) and $1.0$ Gyr (bottom). 
White circles mark the sites of the forming DMDGs. 
The cored ($\gamma=0.1$) halo develops a significantly brighter region, indicating higher $b_T$ values and enhanced star formation efficiency at the DMDG site. 
The cuspy ($\gamma=1$) halo, while forms four DMDGs, has substantially lower $b_T$ values.
}
\label{fig:BKE20_compare}
\end{figure*}

Our simulations reveal a systematic trend in which star formation is more efficient when the gravitational binding is weakened. To demonstrate this, we compare the quantity $b_T$ for the $\gamma=1$ (left) and $\gamma=0.1$ (right) cases in \autoref{fig:BKE20_compare}
This parameter, motivated by the ratio of dynamical time to free-fall time, increases with star formation efficiency. 
Following \citet{2016ApJ...831...16L}, we define
\begin{equation}
b_T \equiv \frac{\Sigma_{\rm gas}}{\sigma_{v,\rm gas}^2 + c_s^2} \propto \left(\frac{\tau_{\rm dyn}}{\tau_{\rm ff}}\right)^2,
\end{equation}
with $c_s = \sqrt{T k_B/(\mu m_p)}$ and $\mu=2.3$.

\begin{figure}[ht!]
\centering
\includegraphics[width=0.5\textwidth]{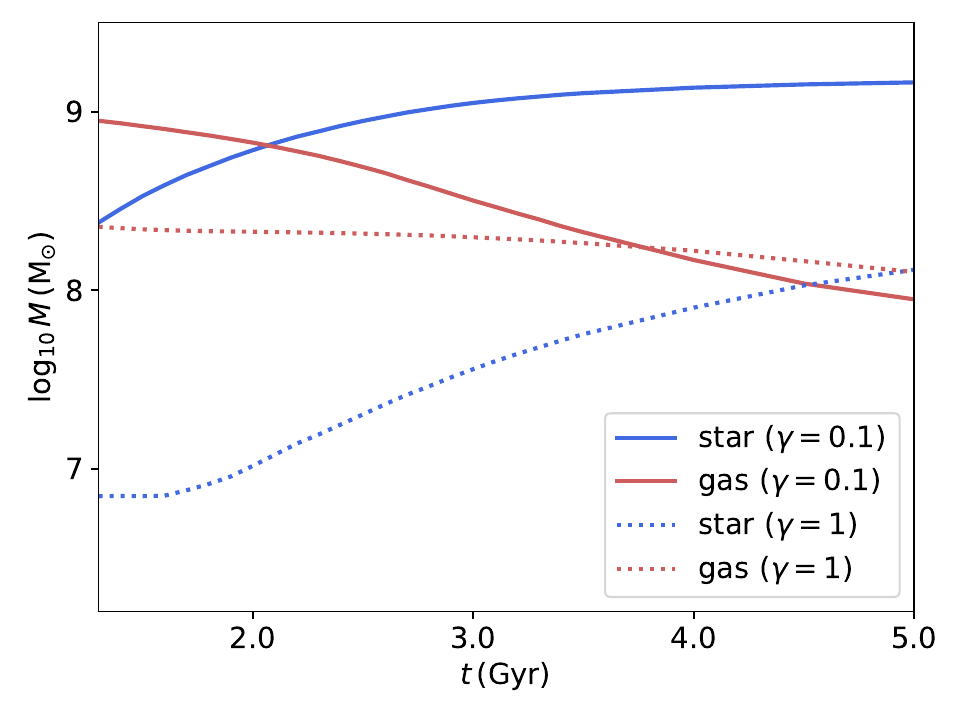}
\caption{\small Mass evolution of the most massive DMDGs in the 2a ($\gamma=1$) and 2b ($\gamma=0.1$) simulations. In the weakened binding case ($\gamma=0.1$), efficient star formation yields a more massive DMDG after the collision, with continued activity beyond 2~Gyr steadily depleting the gas reservoir. 
For $\gamma=1$, star formation is much less efficient, leaving a substantial gas fraction even at 5~Gyr. In both scenarios, the total DMDG mass increases only modestly, from $11\times10^{8}~\rm M_{\odot}$ ($2.3\times10^{8}~\rm M_{\odot}$) to $15\times10^{8}~\rm M_{\odot}$ ($2.6\times10^{8}~\rm M_{\odot}$) in the $\gamma=0.1$ ($\gamma=1$) case.
}
\label{fig:BKE20_mass}
\end{figure}

\autoref{fig:BKE20_compare} shows $b_T$ for the 2a benchmark ($\gamma=1$, left panels) and the 2b benchmark ($\gamma=0.1$, right panels) at $t=0.4$ Gyr (top) and $1.0$ Gyr (bottom). The eventual locations of DMDG formation are enclosed by white circles. In the central collision region, the $\gamma=0.1$ case exhibits systematically higher $b_T$, corresponding to enhanced star formation. Moreover, in the $\gamma=1$ run the gas remnants fragment into three DMDGs, whereas in the $\gamma=0.1$ run only a single remnant forms. Similar behavior is observed across multiple paired simulations.

\autoref{fig:BKE20_mass} compares the mass evolution of the most massive DMDGs in the 2a ($\gamma=1$) and 2b ($\gamma=0.1$) simulations.  
In the weakened gravitational binding scenario ($\gamma=0.1$), efficient star formation produces a larger DMDG mass after the collision. Continued star formation beyond 2~Gyr steadily depletes the gas, leading to a decreasing gas fraction.  
In the $\gamma=1$ case, star formation also reduces the gas fraction, but with much lower efficiency, leaving a substantial gas reservoir even at 5~Gyr.  
In both cases, the total DMDG mass changes only modestly, increasing from $11\times10^{8}~\rm M_{\odot}$ ($2.3\times10^{8}~\rm M_{\odot}$) to $15\times10^{8}~\rm M_{\odot}$ ($2.6\times10^{8}~\rm M_{\odot}$) in the $\gamma=0.1$ ($\gamma=1$) simulation.

\section{Tabulated full simulation results}\label{app:4}

We adopt the gas cooling and star formation module provided in the Gadget-4 code. Radiative cooling is implemented through tabulated rates for primordial gas, while stars form stochastically from dense gas following the multiphase ISM prescription of \citet{Springel:2002uv}.
Although our study does not focus on the detailed dynamical processes within UDGs, the efficiency of gas cooling plays a key role in regulating their stability: a lower cooling rate is generally expected to promote stability. To demonstrate that our conclusions are robust against variations in the cooling rate, we present the full simulation results in \autoref{Tab:2} for a case with the cooling rate reduced by half. In these tables, we identify DMDGs as systems with $M_*(r<10  {\rm kpc}) > 10^{6} M_{\odot}$, reporting their abundance as well as the mass content of the most massive DMDGs in each simulation.

\begin{figure}[ht!]
\centering
\includegraphics[width=0.5\textwidth]{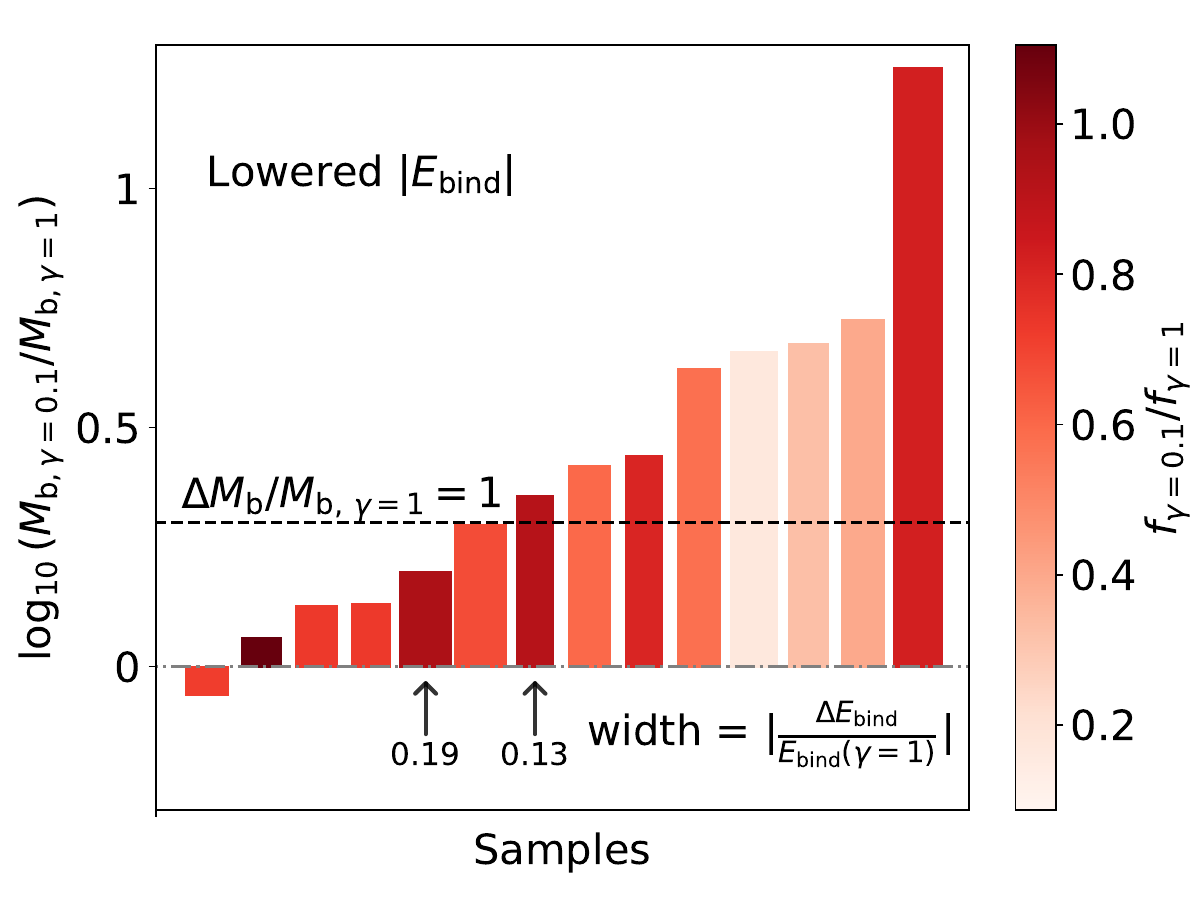}
\caption{\small
Characteristics of the most massive DMDGs in 14 paired simulations with reduced gas cooling. 
The bar chart summarizes results from contrasting simulations with inner slopes $\gamma=0.1$ and $\gamma=1$. Bars are arranged in ascending order of the DMDG mass ratio, $M_{b,\gamma=0.1}/M_{b,\gamma=1}$. The bar width scales with the relative change in binding energy, $(E_{{\rm bind},\gamma=1}-E_{{\rm bind},\gamma=0.1})/E_{{\rm bind},\gamma=1}$, while the color indicates the gas fraction, $f = M_{\rm gas}/(M_{\rm gas}+M_{\rm stars})$. 
In contrast, the increase in the DMDG masses is predominantly positive. In 8 out of the 15 simulations, this increase exceeds 100\%, as reflected by the bars that rise above the line representing $\Delta M_b/M_{b,\gamma=1}=1$. 
}
\label{fig:rc-bars}
\end{figure}

\autoref{fig:rc-bars} summarizes the reduced cooling case, in the same format as \autoref{fig:bars} in the main text. All key features, most notably the enhancement in DMDG mass and star formation efficiency, remain unaffected. 
A comparison of the numerical results across the two tables shows that reduced cooling produces more DMDGs with lower characteristic masses. This indicates a potential degeneracy between the effects of gas cooling and weakened gravitational binding. 
The observed abundance of UDGs may help to distinguish such differences, providing a means to constrain gas cooling~\citep{2017MNRAS.470.4231R,2017A&A...607A..79V,2017MNRAS.466L...1D,2018A&A...614A..21J}.

\begin{figure}[ht!]
\centering
\includegraphics[width=0.5\textwidth]{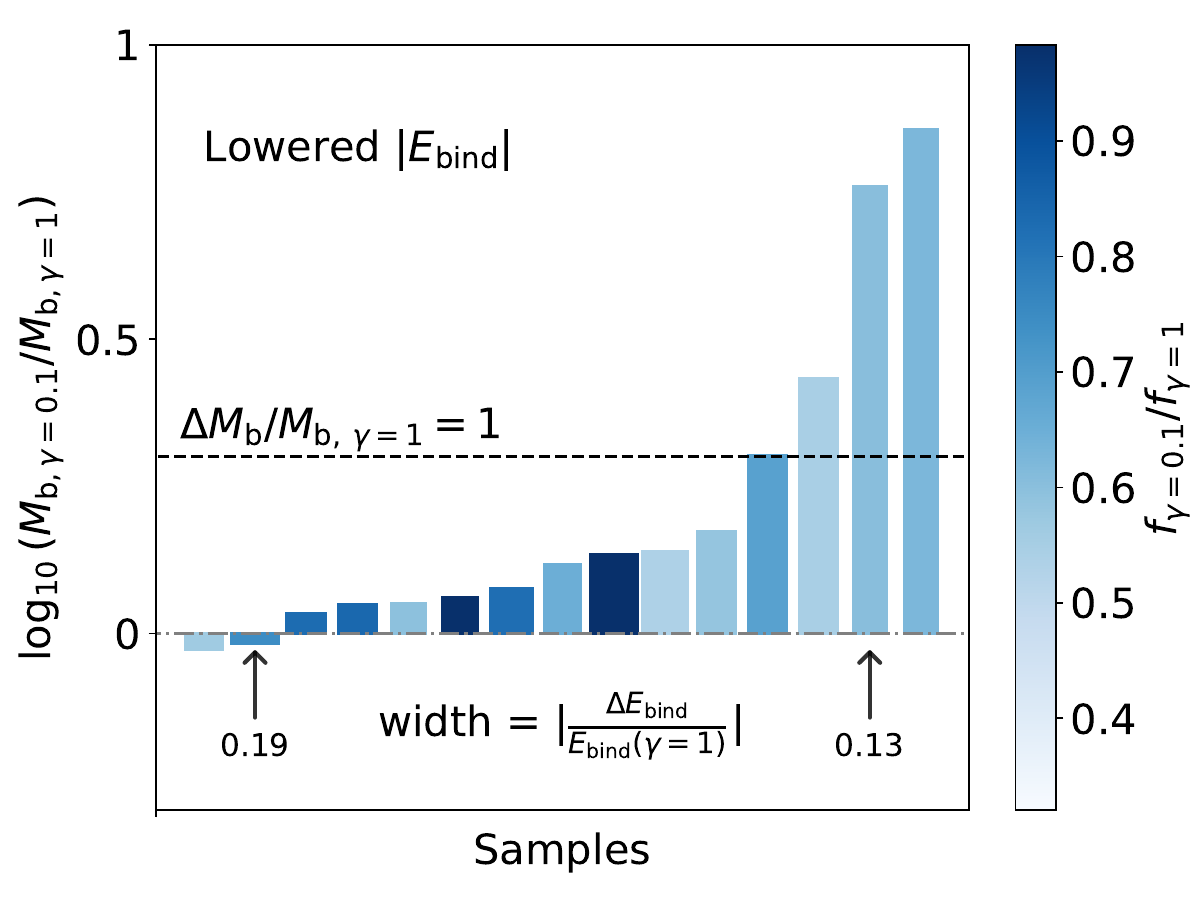}
\caption{\small 
Characteristics of the summed DMDG content for 15 paired simulations. 
The bar chart summarizes results from contrasting simulations with inner slopes $\gamma=0.1$ and $\gamma=1$. Bars are arranged in ascending order of the summed DMDG mass ratio, $M_{b,\gamma=0.1}/M_{b,\gamma=1}$. The bar width scales with the relative change in binding energy, $(E_{{\rm bind},\gamma=1}-E_{{\rm bind},\gamma=0.1})/E_{{\rm bind},\gamma=1}$, while the color indicates the gas fraction, $f = M_{\rm gas}/(M_{\rm gas}+M_{\rm stars})$. 
In contrast, the increase in the DMDG masses is predominantly positive. In 4 out of the 15 simulations, this increase exceeds 100\%, as reflected by the bars that rise above the line representing $\Delta M_b/M_{b,\gamma=1}=1$. 
}
\label{fig:tot-bars}
\end{figure}

Both \autoref{fig:bars} and \autoref{fig:rc-bars} present results for the most massive DMDGs. To further illustrate the robustness of our findings, \autoref{fig:tot-bars} shows the total DMDG masses for the default cooling scenario. In all cases, the trend of enhanced DMDG masses in the weakened gravitational binding scenario remains unchanged.

\begin{deluxetable*}{c|ccccccc|ccc}
\tablecaption{\label{Tab:2}%
Same as Table~\ref{Tab:1}, but with the cooling rate reduced by half. }
\tabletypesize{\footnotesize}
\tablewidth{40mm}  
\tablehead{
\colhead{BM} &\colhead{$M_{200, \, \rm DM}$} & \colhead{$M_{\rm gas}$} & \colhead{$v_r$} & \colhead{$R_s$} & \colhead{$c$} & \colhead{$\gamma$} & \colhead{$|E_{\rm bind}|$} & \colhead{$n$} & \colhead{$M_{\rm star}$} & \colhead{$M_{\rm gas}$} \\
\colhead{} &\colhead{($10^{10}\,M_\odot$)} & \colhead{($10^{10}\,M_\odot$)} & \colhead{(km s$^{-1}$)} & \colhead{(kpc)} & \colhead{} & \colhead{} & \colhead{$E_0$} & \colhead{} & \colhead{($10^{8}\,M_\odot$)} & \colhead{($10^{8}\,M_\odot$)}
}
\startdata
1a & \multirow{2}*{1.5} &\multirow{2}*{0.15} &\multirow{2}*{$400$} &\multirow{2}*{$2$} &\multirow{2}*{$14$} &1 &8.75 &$2$ &$0.02$ &$1.14$\\
1b & & & & & &0.1 &7.58 &$3$ &$0.29$ &$2.35$\\
\hline
2a & \multirow{2}*{2} &\multirow{2}*{0.2} &\multirow{2}*{$400$} &\multirow{2}*{$2$} &\multirow{2}*{$14$} &1 &14.73 &$5$ &$0.02$ &$1.02$\\
2b & & & & & &0.1 &12.49 &$3$ &$2.08$ &$2.28$\\
\hline
3a & \multirow{2}*{2} &\multirow{2}*{0.2} &\multirow{2}*{$280$} &\multirow{2}*{$2$} &\multirow{2}*{$14$} & 1   & 14.73 & 4 & 3.43 & 2.97 \\
3b & &     &     &   &    & 0.1 & 12.49 & 2 & 12.49 & 4.35 \\
\hline
4a & \multirow{2}*{1.0} &\multirow{2}*{0.1} &\multirow{2}*{$400$} &\multirow{2}*{$1.5$} &\multirow{2}*{$4$} &1 &3.53 &$2$ &$0.06$ &$0.55$\\
4b & & & & & &0.1 &2.87 &$3$ &$0.15$ &$0.81$\\
\hline
5a & \multirow{2}*{1.5} &\multirow{2}*{0.15} &\multirow{2}*{$400$} &\multirow{2}*{$2$} &\multirow{2}*{$4$} &1 &7.17 &$2$ &$0.35$ &$2.29$\\
5b & & & & & &0.1 &5.82 &$2$ &$2.33$ &$2.90$\\
\hline
6a & \multirow{2}*{1.5} &\multirow{2}*{0.22} &\multirow{2}*{$400$} &\multirow{2}*{$2$} &\multirow{2}*{$4$} &1 &10.99 &$1$ &$10.56$ &$7.62$\\
6b & & & & & &0.1 &9.28 &$2$ &$11.29$ &$4.56$\\
\hline
8a & \multirow{2}*{1.5} &\multirow{2}*{0.15} &\multirow{2}*{$400$} &\multirow{2}*{$2$} &\multirow{2}*{$7$} &1 &7.87 &$2$ &$0.04$ &$0.26$\\
8b & & & & & &0.1 &6.47 &$1$ &$1.64$ &$3.73$\\
\hline
9a & \multirow{2}*{1.5} &\multirow{2}*{0.22} &\multirow{2}*{$400$} &\multirow{2}*{$2$} &\multirow{2}*{$7$} &1 &11.93 &$3$ &$0.49$ &$2.96$\\
9b & & & & & &0.1 &10.21 &$2$ &$12.68$ &$3.64$\\
\hline
10a & \multirow{2}*{1.5} &\multirow{2}*{0.22} &\multirow{2}*{$280$} &\multirow{2}*{$2$} &\multirow{2}*{$7$} & 1   &11.93 & 2 & 21.53 & 5.88 \\
10b & &      &     &   &   & 0.1 &10.21 & 1 & 23.94 & 7.44 \\
\hline
11a & \multirow{2}*{1.5} &\multirow{2}*{0.15} &\multirow{2}*{$280$} &\multirow{2}*{$2$} &\multirow{2}*{$14$} & 1  &8.75 & 2 & 3.67 & 2.79 \\
11b & &      &     &   &    & 0.1 &7.58 & 1 & 11.84 & 5.98 \\
\hline
12a & \multirow{2}*{1.5} &\multirow{2}*{0.22} &\multirow{2}*{$400$} &\multirow{2}*{$2$} &\multirow{2}*{$14$} &1 &13.56 &$4$ &$1.69$ &$3.53$\\
12b & & & & & &0.1 &11.68 &$3$ &$3.72$ &$3.33$\\
\hline
13a & \multirow{2}*{2} &\multirow{2}*{0.2} &\multirow{2}*{$400$} &\multirow{2}*{$2$} &\multirow{2}*{$4$} &1 &11.72 &$3$ &$0.75$ &$3.64$\\
13b & & & & & &0.1 &9.75 &$1$ &$18.53$ &$1.44$\\
\hline
14a & \multirow{2}*{2} &\multirow{2}*{0.2} &\multirow{2}*{$400$} &\multirow{2}*{$2$} &\multirow{2}*{$7$} &1 &12.81 &$2$ &$0.19$ &$1.91$\\
14b & & & & & &0.1 &10.83 &$2$ &$7.72$ &$3.42$\\
\hline
15a & \multirow{2}*{5} &\multirow{2}*{0.5} &\multirow{2}*{$500$} &\multirow{2}*{$3.5$} &\multirow{2}*{$14$} &1 &78.10 &$5$ &$16.31$ &$4.45$\\
15b & & & & & &0.1 &66.40 &$4$ &$23.62$ &$4.14$\\
\enddata
\end{deluxetable*}

\bibliography{Refs}{}
\bibliographystyle{aasjournalv7}

\end{document}